\documentclass[floats,floatfix,prl,amssymb,twocolumn,superscriptaddress,nofootinbib,10pt]{revtex4-1}

\usepackage{amssymb,amsmath,verbatim,mathtools,needspace,enumitem,etoolbox,graphicx,physics,microtype,afterpage,bm}
\usepackage[dvipsnames, usenames]{xcolor}
\definecolor{linkcolor}{rgb}{0.0,0.3,0.5}
\usepackage[unicode, colorlinks=true, linkcolor=linkcolor, citecolor=linkcolor, filecolor=linkcolor,urlcolor=linkcolor, pdfusetitle]{hyperref}
\usepackage[all]{hypcap}
\usepackage[T1]{fontenc}
\usepackage[utf8]{inputenc}
\usepackage{tabularx}
\usepackage{float}
\usepackage{lmodern}
\allowdisplaybreaks
\usepackage{tikz}
\usepackage{color}
\usepackage{framed}
\usepackage{hyperref}
\hypersetup{colorlinks, citecolor=bluscuro, linkcolor=black, urlcolor=bluscuro}
\definecolor{rossos}{cmyk}{0,1,1,0.55}
\definecolor{bluscuro}{rgb}{0.15, 0.2, .85}
\definecolor{bluchiaro}{cmyk}{1,.3,0.,0.1}
\definecolor{ForestGreen}{rgb}{0.13, 0.55, 0.13}

\newcommand{\be}{\begin{equation}}
\newcommand{\ee}{\end{equation}}
\newcommand{\bea}{\begin{equation}\begin{aligned}}
\newcommand{\eea}{\end{aligned}\end{equation}}

\def\lsim{\mathrel{\rlap{\lower4pt\hbox{\hskip0.5pt$\sim$}}
    \raise1pt\hbox{$<$}}}         
\def\gsim{\mathrel{\rlap{\lower4pt\hbox{\hskip0.5pt$\sim$}}
    \raise1pt\hbox{$>$}}}         

\def\R{{\cal R}}

\newcommand{\nc}{\newcommand}
\nc{\ba}{\begin{eqnarray}}
\nc{\ea}{\end{eqnarray}}
\newcommand{\calR}{{\cal{R}}}
\newcommand{\calP}{{\cal{P}}}

\newcommand{\arXiv}[2]{\href{http://arxiv.org/pdf/#1}{{\tt [#2/#1]}}}
\newcommand{\arXivold}[1]{\href{http://arxiv.org/pdf/#1}{{\tt [#1]}}}

\begin{document}

\title{The Sign of non-Gaussianity  and the Primordial Black Holes Abundance 
}

\author{Hassan Firouzjahi}
\email{firouz@ipm.ir}
\affiliation{School of Astronomy, Institute for Research in Fundamental Sciences (IPM) \\ P.~O.~Box 19395-5531, Tehran, Iran}

\author{Antonio Riotto}
\email{antonio.riotto@unige.ch}
\affiliation{D\'epartement de Physique Th\'eorique, Universit\'e de Gen\`eve, 24 quai E. Ansermet, CH-1211 Geneva, Switzerland}
\affiliation{ Gravitational Wave Science Center (GWSC), Universit\'e de Gen\`eve, 24 quai E. Ansermet, CH-1211 Geneva, Switzerland}


\begin{abstract}
\noindent
The abundance of primordial black holes changes in the presence of local non-Gaussianity. A positive non-linear parameter $f_{NL}$ increases the abundance while a negative one reduces it. We show that in non-attractor  single-field models of inflation which enhance the curvature power spectrum and may give rise to primordial black holes, $f_{NL}$ is always positive,  when computed in correspondence of the peak of the curvature power spectrum where the primordial black hole abundance has its maximum.  This implies that the interpretation of the recent pulsar timing arrays data from scalar-induced gravitational waves generated at primordial black hole formation may not be supported  by invoking   non-Gaussianity  within non-attractor single-field models.
\end{abstract}

\maketitle

\noindent{{\bf{\it Introduction.}}}
Very recently the  NANOGrav~\cite{NANOGrav:2023gor,NANOGrav:2023hde}, EPTA~\cite{EPTA:2023fyk,EPTA:2023sfo,EPTA:2023xxk}, PPTA~\cite{Reardon:2023gzh,Zic:2023gta,Reardon:2023zen} and CPTA~\cite{Xu:2023wog} collaborations have provided  evidence for a stochastic background of Gravitational Waves (GWs) detected through the pulsar timing arrays. One immediate question is under which circumstances such GWs can be associated to the formation of Primordial Black Holes (PBHs) during which GWs are inevitably generated at second-order \cite{Sasaki:2018dmp}.  Their amount is proportional to the the square of the amplitude of the dimensionless curvature perturbation power spectrum ${\cal P}_\R$, $\Omega_{\text{\tiny GW}}\sim {\cal P}_\R^2$. The abundance
of PBHs is exponentially  sensitive to the same amplitude,  $f_{\text{\tiny PBH}}\sim {\rm exp}(-1/{\cal P}_\R)$, where $f_{\text{\tiny PBH}}$ is the PBH abundance with respect to the total dark matter. The problem is that the observed stochastic GW background is  explained   by a  relatively large values of ${\cal P}_\R$,  which has been claimed to  lead to a too large PBH abundance \cite{Franciolini:2023pbf,Liu:2023ymk,Cai:2023dls,Inomata:2023zup,Zhu:2023faa}. While this negative conclusion may be  invalidated by the recent observation that  corrections  from the non-linear radiation transfer function and the determination of the true physical horizon crossing  decrease the PBH abundance
\cite{DeLuca:2023tun}, one can also rely on the introduction of some local Non-Gaussianity (NG) in the curvature perturbation

\be
\R=\R_{\rm g}+\frac{3}{5}f_{NL}\left(\R^2_{\rm g}-\langle\R^2_{\rm g}\rangle\right),
\ee
where $\R_{\rm g}$ is the Gaussian component\footnote{We are adopting this quadratic expansion to be model independent, even though in general the exact  relation between $\R$ and $\R_{\rm g}$ can be worked out model by model. However, since typically $f_{NL}\R_{\rm g}\lsim 1$, the quadratic expansion  is justified.}. 
The short-scale power spectrum  ${\cal P}_S$ responsible for the PBH formation is modulated by the presence of a long mode $\R_L$. The threshold $\R_c$ for the formation of the PBHs is shifted approximately by  \cite{Young:2015kda}

\be
\R_c\simeq \R_c^{\rm g}\left(1-\frac{3}{5}f_{NL}\R_c^{\rm g}\right),
\ee
compared to the threshold  $\R_c^{\rm g}$ in the Gaussian theory.
Therefore, around peaks of the power spectrum of the curvature perturbation,  a positive $f_{NL}$ increases the abundance of the PBHs, while a negative $f_{NL}$ has the opposite effect, thus helping the agreement with the recent pulsar timing array observations. This remains true even  when calculating the abundance through a more  correct variable, the averaged density contrast \cite{Kehagias:2019eil,DeLuca:2022rfz}.

Under general assumptions, in this paper  we will show that the sign of $f_{NL}$ at the peak scale of the power spectrum, where PBHs are mostly formed, is 
always positive   in  non-attractor  single-field models.  This no-go  result  is intimately  related to the fact that $f_{NL}$ measures the response of the short-scale power spectrum ${\cal P}_S$ to the presence of a long mode and the sign of the NG is determined by the rate of growth of ${\cal P}_S$. The latter is positive if PBHs needs to be produced and this sets the sign of $f_{NL}$. Our findings  automatically imply that NG may  not help non-attractor  single-field models to relax the tension between the observed stochastic GW background in pulsar timing arrays and the overproduction of PBHs.

\vskip 0.5cm
\noindent{{\bf{\it Non-attractor single-field models and the sign of NG.}}} In attractor single-field models the curvature perturbation is constant on superhorizon scales and is equivalent in the spatially flat gauge to a field fluctuation $\R=-\delta\phi/\phi'$, where primes denote derivatives with respect to the number of e-folds. The phase-space trajectory of the long mode
perturbation follows that of the background itself.
Short-scale modes evolving  in a long mode 
perturbation then  follow the phase-space trajectory
of the background, with the only difference being the
local e-folds which determines the relation between
the  comoving and the physical wavenumbers. The NG is therefore proportional to the variation of the short-scale power spectrum due to the long-wavelength mode

\begin{eqnarray}
{\cal P}_S(x)&=&{\cal P}_S\left[1-\frac{d\ln {\cal P}_S}{d\ln k_S}\R_L(x)\right]\nonumber\\
&=&\left[1+\frac{12}{5}f_{NL}\R_L(x)\right].
\end{eqnarray}
This modulation is zero at the peak of the short-scale power spectrum and corresponds to a dilation of scales
rather than an amplitude enhancement. 

In non-attractor  single-field models, the attractor condition $\delta\phi'=(\phi''/\phi')\delta\phi$ is violated. In fact, during an Ultra-Slow-Roll (USR) phase, the curvature perturbation grows like $\R\sim a^3$, being $a$ the scale factor, and therefore in the spatially flat gauge $\delta\phi=-\phi'\R=$ constant, implying that $\delta\phi'=0$. Because of  the
the dependence of the background evolution on the initial kinetic energy,
the perturbation may not  be mapped into a
change in the background clock along the same phase-space
trajectory. The long mode  perturbations carry no corresponding $\delta\phi'$ and so they shift the USR trajectory
to one with a different relationship between $\phi$ 
and $\phi'$. In other words, a  local measurement is sensitive
to $\phi'$ as  different observers provide different measurements of the short-scale power spectrum
depending on their relative position in the long-wavelength
mode. This implies that in USR models the corresponding value of $f_{NL}$ can be large, even at the peak of the short-scale power spectrum.

We consider single field models of inflation with the potential $V(\phi)$
for a canonically normalized scalar field with the sound speed of perturbations being equal to the speed of the gravitational waves perturbations. 
To be general, we do not specify the form of the potential.
We assume that inflation has multiple stages, containing at least three distinct phases. The first stage  is a conventional slow-roll (SR) phase in which the observed large scales, such as the CMB scales, leave the horizon. The 
power spectrum of  these perturbations are fixed by the CMB observations \cite{Planck} to be $\calP_\calR \simeq 2\times 10^{-9}$ with $\calR$ being the curvature perturbations.
The second phase is when the power spectrum experiences a rapid growth with a prime  peak in power spectrum to generate PBHs \cite{Byrnes:2018txb, Cole:2022xqc, Carrilho:2019oqg, t, Ozsoy:2021pws}.   
A common mechanism for the enhancement of the power spectrum may be the USR setup where the potential is flat  \cite{Kinney:2005vj, Namjoo:2012aa}. However, we consider a general case and for this purpose, we may call this intermediate non-attractor phase as a ``USR-type'' phase. All we require from the form of the potential to be such that 
the power spectrum to increase  monotonically  during the second phase. The final phase is an attractor SR regime which is extended towards  the end of inflation. The transitions between the stages  can be either sharp or mild. 
We present our results for a three-phase setup SR $\rightarrow $ non-attractor $\rightarrow $ SR, and the extension of the results to higher  multiple phases  is 
straightforward. 
We do not consider the stochastic random motion of the background field so the behaviour of $\phi$ is monotonic. 
The  non-attractor  phase is extended in the region $\phi_e < \phi< \phi_s$ during the time interval $t_s < t< t_e$
and 
we are interested in the growth of power spectrum for the modes which leave the Hubble radius during the non-attractor phase.

For PBH formation, we are interested in the short-scale power spectrum and in particular the PBH mass function will be dominated by the PBHs forming when the scale  $k_{\rm pk}$ corresponding to  the peak of the power spectrum will re-enter the Hubble radius. Let us consider therefore the effect of the long mode $k_L\lsim k_{\rm pk}\sim k_S$. Notice that  long mode is itself suffering a  period of USR phase, but it has exited the Hubble radius earlier than the scale  $k_S$. The measurements of the power spectrum and the bispectrum are made at the end of inflation $t=t_f$ when the modes are frozen. The effects of the long mode on the short modes can be viewed as the modulation of the background quantities at the
end of non-attractor phase $t=t_e$.  As in separate universe approach, one can view the effects of the long mode as affecting nearby patches slightly differently.  Consequently,  different patches approach the final attractor phase with slightly different initial conditions modulated by the long mode at the end of non-attractor phase.  With this picture in mind the bispectrum for two short modes 
under the modulation of a long mode can be written as 
\begin{eqnarray}
\label{modulation0}
\Big<\R^{f}_L \R^{f}_S \R^f_S \Big>\simeq \Big< \R^f_L\,  \Big< \calR^f_S  \calR^f_S\Big>_{\calR^e_L}  \Big> 
\end{eqnarray}
in which $\calR_S$ and $\calR_L$ represent the short and long modes while the superscript $f$ and $e$ indicate the corresponding values at $t=t_f$ and $t=t_e$,
respectively. The assumption of having a  single-field setup is essential in writing the above relation. If there are extra light fields, then one has to include the modulations by them  in the right-hand side of Eq. (\ref{modulation0}) as well.

In non-attractor single-field models $\R_L$ and $\dot\R_L$ are to be  treated as independent variables \cite{Namjoo:2013fka}. 
Expanding $ \big< \calR^f_S  \calR^f_S\big>_{\calR^e_L} $ to leading order  yields  
\begin{eqnarray}
\label{modulation}
\Big<\R^f_L \R^f_S\R^f_S\Big>&\simeq& \Big< \R^f_L\left(\R^e_L\frac{\partial}{\partial\R^e_L}\langle \R^f_S\R^f_S\rangle\right.\nonumber\\
&+&\left. \dot\R^e_L\frac{\partial}{\partial\dot \R^e_L}\langle \R^f_S\R^f_S\rangle\right)\Big> .
\end{eqnarray}
An implicit assumption in performing the above expansion is that $\calR$ and
$\dot \calR$ to be continuous across the transition. This is the usual assumption that 
one needs to impose for the continuity of the metric and the extrinsic curvature across the transition. Having said this, we do not impose any assumption 
 on the potential $V(\phi)$ and its derivatives, as long as $\calR$ and
$\dot \calR$ are continuous across the transition.

Expressing the left hand side of Eq. (\ref{modulation}) in terms of the usual 
non-Gaussianity parameter $f_{NL}$ and defining the power spectrum in Fourier space as $\big \langle \calR_{\bf k_1} \calR_{\bf k_2} \big \rangle= (2 \pi)^3 \delta^3({\bf{k_1} }+ {\bf{k_2}}) P(k_1)$ and discarding the trivial factors of
$(2 \pi)^3 \delta^3(\bf{k})$ which matches automatically from the momentum conservation, we obtain  
\begin{eqnarray}
\label{fNL1}
\frac{12}{5}f_{NL}{ P}^f_L{ P}^f_S&\simeq &  \langle \calR^f_L \calR^e_L \big \rangle
\frac{\partial { P}^f_S}{\partial\R^e_L} + 
\big \langle \calR^f_L \dot \calR^e_L \big \rangle
\frac{\partial { P}^f_S}{\partial\dot \R^e_L},
\end{eqnarray}
in which $P^f_S$ and $P^f_L$ represents the power spectrum at the end of inflation for the short and long modes respectively.

From the above expression we have to calculate correlations like
$\big \langle \calR^f_L \calR^e_L \big \rangle$  for the long mode perturbations
at two different times $t_e$ and $t_f$. 
As explained before, this is because the long mode at the end of non-attractor phase modulates the power spectrum of the short modes which are measured at the end of inflation. 

Since the long mode is far outside the horizon at the end of non-attractor phase, we can treat it as classical and  relate $\big \langle \calR^f_L \calR^e_L \big \rangle$ to $P^f_L$ via the ratio of the mode functions at these two times:
\ba
\big \langle \calR^f_L \calR^e_L \big \rangle= \left(\frac{\calR^e_L}{\calR^f_L}\right) P^f_L,
\ea
and similarly
\ba
\big \langle \calR^f_L \dot \calR^e_L \big \rangle= \frac{1}{2}\left(\frac{\calR^f_L}{\calR^e_L}\right) \frac{dP^e_L}{d t}.
\ea
Plugging the above relations into Eq. (\ref{fNL1}) yields 
\begin{eqnarray}
\label{fNL2}
\frac{12}{5}f_{NL} &=& \left(\frac{\R^e_L}{\R^f_L}\right) \frac{\partial \ln \calP^f_S}{\partial \calR^e_L}
 + \frac{1}{2} \left(\frac{\R^f_L}{\R^e_L}\right) \frac{ \dot \calP^e_L }{\calP^f_L} 
 \frac{\partial \ln \calP^f_S}{\partial \dot \calR^e_L},\nonumber\\
 &&
\end{eqnarray}
in which the dimensionless power spectrum $\calP_\calR$ is related to the power spectrum via 

\be
\calP_\calR \equiv \frac{k^3}{2 \pi^2} P_\calR.
\ee 
We should now trade the two independent variables $(\calR_L, \dot\calR_L)$ with two other variables in which then partial derivative has a more transparent  meaning.
From the point of view of a local observer within a region of size $\sim 1/k_S$, the long mode perturbation evolves with time, but with negligible spatial gradients so the metric takes the following form

\ba
ds^2= - dt^2 + a^2(t) e^{2 \calR_L(t)}d {\bf x}^2,
\ea
We can absorb the long mode into the scale factor via
$\widetilde a\equiv a e^{\calR_L}$ and the corresponding Hubble rate will change as $\widetilde H = H + \dot \calR_L$. Consequently 
\ba
\label{da}
d \ln \widetilde a = d \calR_L
\ea
and 
\ba
\label{dH}
d\widetilde H=  d \dot \calR_L.
\ea
Eqs. (\ref{da}) and (\ref{dH}) are two differential relations that can be used to relate 
$(d \calR, d \dot \calR)$ to $(d \ln \widetilde a, d \ln \widetilde H)$. More specifically, we have
\ba
\label{diff1}
d \ln \calP_S &=& \frac{\partial \ln{ \calP}_S}{\partial\R_L} d \calR_L + 
\frac{\partial \ln{ \calP}_S}{\partial \dot \R_L} d \dot \calR_L \nonumber\\
&=&\frac{\partial \ln{ \calP}_S}{\partial \ln \widetilde a} d \ln \widetilde a + 
\frac{\partial \ln{ \calP}_S}{\partial  \widetilde H } d  \widetilde H.
\ea
Using the relations between  $(d \ln \widetilde a, d  \widetilde H)$ and $(d \calR, d \dot \calR)$, from the second line of the above equation  
we obtain
\ba
\label{diff2}
d \ln \calP_S =    \frac{\partial \ln{\calP}_S}{\partial  \ln \widetilde a} d \calR_L +  \frac{\partial \ln{ \calP}_S}{\partial  \widetilde H }  {d \dot \calR_L}.
\ea
Comparing this differential equation with the first line of Eq. (\ref{diff1}) we obtain
\ba
\label{eq1}
 \frac{\partial \ln{ \calP}_S}{\partial\R_L}=  \frac{\partial \ln{ \calP}_S}{\partial \ln \widetilde a}
\ea
and
\ba
\label{eq2}
\frac{\partial \ln{ \calP}_S}{\partial \dot \R_L}=   \frac{\partial \ln{ \calP}_S}{\partial  \widetilde H }.
\ea
Now, plugging the above relations into formula (\ref{fNL2}) and replacing 
$\widetilde a $ and $\tilde H$ simply by $a$ and $H$ 
yields
\begin{eqnarray}
\label{fNL3a}
\frac{12}{5}f_{NL}&=&  \left(\frac{\calR^e_L}{\calR^f_L}\right) \frac{\partial \ln{\calP}^f_S}{\partial \ln  a_e} + \left(\frac{\calR^f_L}{\calR^e_L}\right) \frac{\dot \calP^e_L}{2 H^2_e\calP^f_L} \frac{\partial \ln{\cal P}^f_S}{\partial  \ln H_e }.\nonumber\\
&&
\end{eqnarray}
One can think of Eq. (\ref{fNL3a})
as an extension of Maldacena's consistency condition \cite{Maldacena:2002vr} to the non-attractor setups (see also \cite{cr1,cr2}). The importance of this consistency condition is that we can read off the value of $f_{NL}$ from the properties of the power spectrum and without the need to calculate the bispectrum using either $\delta N$ or in-in formalisms for higher orders perturbation theory.

So far our analysis was general relying only on the assumption of a single-field 
inflation model undergoing non-attractor phase(s) during inflation. 
The working assumption is that the power spectrum experiences rapid growth until it reaching a peak associated to the narrow scale where PBHs are formed. For the modes which leave the Hubble radius during the non-attractor phase and near the peak,  the power spectrum locally has the following form in momentum space 
\ba
\label{power-app}
\calP_S =f(a_e) \left( \frac{k_S}{a_e H_e} \right)^{n_\calR-1},
\ea 
in which $n_\calR$ is the spectral index and $f(a)$ is a function of the background 
which controls the rapid growth of the power spectrum. Technically speaking,
the factor $f(a)$ comes from the fact that the first slow-roll parameter 
$\epsilon \equiv -\dot H/H^2$ falls off rapidly during the non-attractor phase so the 
the power spectrum $\calP \propto \epsilon^{-1}$  experiences a rapid growth 
during the non-attractor phase. For example, in the conventional USR phase
$\epsilon \propto a^{-6}$ and correspondingly $f(a) = a^6$. In our analysis, we do not rely on the particular type of the transition and the 
form of $f(a)$ and all we assume is that $f(a)$ is a growing function of $a$ to ensure the rapid growth of $\calP_\calR$ during the non-attractor phase. We emphasize again that the form of power spectrum given in Eq. (\ref{power-app}) is valid only locally near the peak which is followed by a rapid increase in power spectrum. The general form of the power spectrum in $k$-space is more complicated and may not be even described by a power law behaviour. For example, it can have oscillatory features after the prime peak as in conventional USR setup \cite{Byrnes:2018txb, Cole:2022xqc, Carrilho:2019oqg, t, Ozsoy:2021pws}. However, since we are interested in power spectrum slightly prior and around the peak associated to the narrow scales where the PBHs are formed, then the ansatz (\ref{power-app}) is physically justified. 

From Eq. (\ref{power-app}) we infer

\ba
\frac{\partial \ln{\cal P}^f_S}{\partial  \ln H_e }= -\frac{d\ln{\cal P}^f_S}{d\ln k_S }=1- n_\R.
\ea
Near the peak of the power spectrum by definition $(n_\R-1)\simeq 0$ and correspondingly we obtain 
\ba
\label{fNL3}
f_{NL}^{\rm pk}=   \frac{5}{12}  \left(\frac{\calR^e_L}{\calR^f_L}\right)     \frac{\partial  \ln{\cal P}^f_S}{\partial \ln a_e}.
\ea
We note that the prefactor $(\calR^e_L/\calR^f_L)$ appears 
because the mode function in general evolves after the non-attractor phase. This is because the transition from the non-attractor phase to the final attractor phase may be mild so the mode keeps evolving in time until it reaches 
its final attractor value \cite{c}. The long mode is far outside the horizon after the peak, evolving from its initial value  $\calR^e_L$ at $t=t_e$ to its final value $\calR^f_L$  at $t=t_f$. Therefore,  $\calR^f_L$ is in phase with $\calR^e_L$ in $k$-space. However, as the background quantities such as the slow-roll parameters are evolving during a mild transition, the mode function may change sign so the ratio 
$(\calR^e_L/\calR^f_L)$  may become negative. On the other hand, if the transition is mild,  then the peak in power spectrum will not be significant as the power spectrum evolves in subsequent evolution so it is not a viable model for PBHs formation in the first place. Therefore, in what follows, we make an implicit assumption that  the transition from the intermediate non-attractor phase 
 to the final attractor phase is sharp enough such that $(\calR^e_L/\calR^f_L)$ remains positive.   Since the power spectrum is an increasing function of time during the intermediate non-attractor phase, we conclude that 
\be
f_{NL}^{\rm pk}>0.
\ee

While our conclusion about the sign of $f_{NL}^{\rm pk}$ is general (with the implicit assumption of a sharp enough transition), let us examine it for some non-trivial examples. Let us consider a setup in which a USR phase is followed by an attractor SR phase in which the transition to the final attractor phase 
can be either sharp or mild.  Defining the slow-roll parameter associated to the derivative of the potential at the final attractor phase by $\sqrt{2 \epsilon_V} \equiv V_\phi/V$, the sharpness of the transition from the intermediate USR phase to the final attractor phase is determined by the parameter $h$ given by \cite{c}
\ba
\label{h-def}
h\equiv  -6 \sqrt{ \frac{\epsilon_V} {\epsilon_e}} \, ,
\ea
in which $\epsilon_e$ is the value of the slow-roll parameter at the end of USR phase. Note that in this convention $h<0$. 
For a very sharp transition $| h| \gg 1$ while for a mild transition $h $ may be comparable to slow-roll parameters. In order to have sharp enough transition such that  the ratio $(\calR^e_L/\calR^f_L)$ remains positive, we assume $\eta_V\rightarrow 0$ in which $\eta_V$ is the second slow-roll parameter given by 
$\eta_V = V_{\phi \phi}/V$.    

The mode function  for the modes which leave the horizon during the USR phase is given by \cite{c}
\ba
\label{mode1}
\calR^f_k = \left( 1+ \sqrt{ \frac{\epsilon_V} {\epsilon_e}}  \right)
\frac{H}{\sqrt{4 \epsilon_V k^3}}.
\ea
Since during the USR phase the slow-roll parameter falls off like 
$a^{-6}$, then  $\epsilon_e \propto a_e^{-6}$.  Taking the derivative 
with respect to $a_e$ we find
\ba
\label{diff-P}
 \frac{d \ln \calP^f}{d \ln a_e} = 6 \sqrt{ \frac{\epsilon_V} {\epsilon_e}} \left( 1+ \sqrt{ \frac{\epsilon_V} {\epsilon_e}}  \right)^{-1}  
 = \frac{6h}{h-6} .
\ea
On the other hand, the ratio $\calR^e_L/\calR^f_L$ yields an additional factor
\ba
\label{ratio}
\frac{\calR^e_L}{\calR^f_L}= \sqrt{ \frac{\epsilon_V} {\epsilon_e}} \left( 1+ \sqrt{ \frac{\epsilon_V} {\epsilon_e}}  \right)^{-1} 
=  \frac{h}{h-6}  . 
\ea
We see that the ratio $(\calR^e_L/\calR^f_L)$ is positive as expected. 
Using Eqs. (\ref{diff-P}) and (\ref{ratio}) in our formula (\ref{fNL3}) yields 
\ba
\label{fNL-h}
f_{NL}^{\rm pk}= \frac{5 h^2}{2 (6-h)^2}>0. 
\ea
For an infinitely sharp transition with $h\rightarrow -\infty$ in which  the mode function is frozen immediately after the transition with $\calR^e_L= \calR^f_L$, from Eq. (\ref{fNL-h}) we obtain 
the expected result $f_{NL}^{\rm pk}=5/2$. The expression Eq. (\ref{fNL-h})
agrees with the result for $f_{NL}$ obtained in  \cite{c} where the power spectrum is  scale-invariant as well. 

As a second example, now suppose we extend the above setup such that there is an upward shift $\Delta V$ in the potential at the end of non-attractor phase, followed by the final SR phase. As in Ref.  \cite{Cai:2022erk}, suppose the upward step in the potential is instantaneous, yielding to a sudden change in  inflaton's velocity. Imposing the conservation of energy,   the inflaton velocity at the end of upward transition  $\pi_d$ is related to the velocity at the end of no-attractor phase $\pi_e$ via
\ba
\pi_d= - \sqrt{\pi_e^2 - 6 \frac{\Delta V}{V}} \, ,
\ea
in which  $\pi \equiv\phi'$ with a prime denoting the derivative with respect to the number of e-folds.  The linear mode function is given by \cite{Cai:2022erk}
\ba
\label{mode2}
\calR^f_k = \left( \frac{1}{g}+ \sqrt{ \frac{\epsilon_V} {\epsilon_e}}  \right)
\frac{H}{\sqrt{4 \epsilon_V k^3}}, 
\ea
in which $g\equiv \pi_d/\pi_e$ with $0< g<1$.  Correspondingly, this yields
\ba
\label{diff-P2}
 \frac{d \ln \calP^f}{d \ln a_e} = \frac{6 h g^4 + 36 g^2 - 36}{g^2 (g^2h - 6)} ,
\ea
in which the sharpness parameter $h$ is now defined as $h \equiv -(6/g)\sqrt{\epsilon_V/\epsilon_e}$. 
In addition, the ratio of the mode functions is given by
\ba
\label{ratio2}
\frac{\calR^e_L}{\calR^f_L}=  \frac{h g^2}{h g^2-6} >0  . 
\ea
Note that if we set $g=1$ so $\Delta V=0$, Eqs. (\ref{ratio2}) and (\ref{diff-P2}) reduce to Eqs. (\ref{ratio}) and (\ref{diff-P}) respectively. 
Now plugging Eqs. (\ref{ratio2}) and (\ref{diff-P2}) into our master formula Eq.
(\ref{fNL3}) yields
\ba
\label{fNL-gh}
f_{NL}^{\rm pk}= \frac{5h ( h g^4 + 6 g^2 - 6)}{ 2(g^2h - 6)^2} ,
\ea
in exact agreement with \cite{Cai:2022erk}  for a scale-invariant power spectrum.  If we set $g=1$, corresponding to no bump in potential, then Eq. (\ref{fNL-gh}) reduces to Eq. (\ref{fNL-h}). Noting that $h<0$ and $0<g<1$,
one can check that $f_{NL}^{\rm pk}>0$ for all allowed values of $(h, g)$  as our theorem predicts.  Note that the above value of $f_{NL}$ was calculated in \cite{Cai:2022erk} using the $\delta N$ formalism to second order in perturbation theory. However, in our approach based on consistency condition, we only need to calculate the linear mode function without the need to go to higher orders in perturbation theory.  \\

As a corollary, our theorem implies that in the setups where  the power spectrum experiences a suppression going through a minimum, then $f_{NL}<0$ at the minimum as was observed in a specific setup in \cite{Domenech:2023dxx}.

\vskip 0.5cm
\noindent
\noindent{{\bf{\it Conclusions.}}}
In this  note we have shown that the non-linear parameter $f_{NL}$ in single-field non-attractor models is always positive if calculated for the peak of the enhanced power spectrum. This result implies the NG always increases the PBH abundance.  The sign of the NG is fixed by the response of the short-scale power spectrum to the presence of a long mode. If PBHs need to be form, the  short-scale power spectrum needs to grow and this set the sign of $f_{NL}^{\rm pk}$ uniquely. This logic implies that 
our no-go result   does not hold in the case in which the NG is generated after the inflationary phase, e.g. in the presence of a spectator field.  Indeed, 
one can generate PBHs within a spiky model where the comoving curvature power spectrum is enhanced at small scales through a spectator  isocurvature field   \cite{Kawasaki:2012wr}. This isocurvature perturbation  will then subsequently decay into radiation perturbation and  become a curvature mode after inflation. In such a case  the  long mode cannot be reabsorbed by a redefinition of the scale factor and therefore the sign of the NG is not defined. As a consequence,  $f_{NL}$ can be negative in models with  extra fields. We comment that our conclusion about the 
sign of $f_{NL}^{\rm pk}$ requires an implicit assumption that the transition from the non-attractor phase to the final attractor phase be sharp enough so the mode function keeps its original sign. Physically, this is the relevant case for PBHs formation since if the transition is not sharp enough, then the peak is not prominent and PBHs may not form in the first place.

\vskip 0.5cm
\noindent
\noindent{{\bf{\it Acknowledgments.}}}
H.F. thanks the Department of Theoretical Physics at the University of Geneva for the kind hospitality when part of this work has been done. We thank M. Sasaki and M. H.  Namjoo for insightful discussions and comments. A.R. thanks the Boninchi Fundation for support.



\end{document}